\documentclass[traditabstract]{aa}
\usepackage{graphicx}
\usepackage{stfloats}                                                   
\usepackage{txfonts}
\usepackage{color}
\newcommand{\tablenotea}[1]{\parbox{ 8.6cm}{\indent \footnotesize{#1}}}

\newcommand{\jms}{J. Mol. Spec.}
\newcommand{\ijms}{Int. J. Mass Spectr.}
\newcommand{\jcscc}{J. Chem. Soc. Chem. Commun.}
\newcommand{\cpl}{Chem. Phys. Lett.}
\newcommand{\ieee}{IEEE Trans. Antennas Propag.}
\newcommand{\pccp}{Phys. Chem. Chem. Phys.}

\begin{document}
\title{New molecules in IRC~+10216: confirmation of C$_5$S and tentative identification of MgCCH, NCCP, and SiH$_3$CN\thanks{Based on observations carried out with the IRAM 30m Telescope. IRAM is supported by INSU/CNRS (France), MPG (Germany) and IGN (Spain).}}
\titlerunning{C$_5$S, MgCCH, NCCP, and SiH$_3$CN in IRC~+10216}
\authorrunning{Ag\'undez et al.}

\author{
Marcelino Ag\'undez\inst{1},
Jos\'e Cernicharo\inst{1}, and
Michel Gu\' elin\inst{2}}

\institute{
Instituto de Ciencia de Materiales de Madrid, CSIC, C/ Sor Juana In\'es de la Cruz 3, 28049 Cantoblanco, Spain \and
Institut de Radioastronomie Millim\'etrique, 300 rue de la Piscine, 38406 St. Martin d'H\'eres, France}

\date{Received; accepted}


\abstract
{The C-star envelope IRC~+10216 harbors a rich variety of molecules, with more than 80 detected to date. During the course of a $\lambda$ 3 mm survey of IRC~+10216 carried out with the IRAM 30-m telescope we have detected various weak lines, with antenna temperatures of a few mK, that we assign to rotational transitions of four new molecules. The observation of three lines of C$_5$S confirms a previous tentative identification of this molecule by Bell et al. (1993) based on a line at 24.0 GHz. We also report the tentative identification of three molecules not yet observed in space: MgCCH, the first metal acetylide detected in space, and NCCP and SiH$_3$CN, the phosphorus and silicon analogs of cyanogen (NCCN) and methyl cyanide (CH$_3$CN). We derive the following column densities: $N$(C$_5$S) = (2-14) $\times$ 10$^{12}$
cm$^{-2}$ (depending on the rotational temperature adopted), $N$(MgCCH) = 2 $\times$ 10$^{12}$ cm$^{-2}$, $N$(NCCP) = 7 $\times$ 10$^{11}$ cm$^{-2}$, and $N$(SiH$_3$CN) = 10$^{12}$ cm$^{-2}$. The S-bearing carbon chain C$_5$S is less abundant than C$_3$S, while MgCCH has an abundance in between that of MgNC and those of MgCN and HMgNC. On the other hand, NCCP and SiH$_3$CN are
the least abundant P- and Si-bearing molecules observed to date in IRC~+10216. Based on the behavior of similar molecules it is likely that these four species are formed in the outer circumstellar layers of IRC~+10216. We discuss possible gas-phase formation routes.}
{}
{}
{}
{}

\keywords{astrochemistry -- stars: circumstellar matter -- stars: AGB and post-AGB -- stars: carbon -- stars: individual: IRC~+10216 -- radio lines: stars}

\maketitle

\section{Introduction}

The C-star envelope IRC~+10216 is one of the richest molecular sources in the sky, with more than 80 molecules discovered to date. The central star being at the tip of the AGB phase, it is expelling matter at a high rate of $\sim$2 $\times$ 10$^{-5}$ M$_{\odot}$ yr$^{-1}$, and being relatively nearby ($\sim$130 pc),
it is the prototype of carbon star. The inner wind is a factory of dust and relatively stable molecules while in its outer layers the photochemistry driven by the penetration of interstellar UV photons produces a wealth of exotic molecules.

IRC~+10216 has been intensively observed from cm- to submm-wavelengths, including various line surveys (see a review in \cite{cer2011} 2011). Because of the carbon-rich character of the star, the molecular inventory of the envelope is dominated by carbon chains such as polyyne and cyanopolyyne radicals (\cite{cer1996} 1996; \cite{gue1998} 1998), carbon chains containing sulfur or silicon (\cite{cer1987} 1987; \cite{ohi1989} 1989; \cite{bel1993} 1993), exotic molecules containing phosphorus
or metals (\cite{hal2008} 2008; \cite{cab2013} 2013), and negatively charged carbon chains (\cite{mcc2006} 2006; \cite{cer2007} 2007; \cite{agu2010} 2010).

During the last decades IRC~+10216 has been widely observed with the IRAM 30-m telescope, covering most of the atmospheric spectral windows available at wavelengths of 3, 2, 1.3, and 0.9 mm. In particular, the line survey in the 3 mm band has reached a very high sensitivity in certain spectral regions, allowing to detect very weak features at the level of a few mK of antenna temperature. Here we report various weak lines observed within the $\lambda$ 3 mm band, three of which are assigned to C$_5$S, confirming a previous tentative identification of this molecule based on a line at cm-wavelengths (\cite{bel1993} 1993), while others are tentatively attributed to the molecules MgCCH, NCCP, and SiH$_3$CN.

\section{Observations}

The observations were made with the IRAM 30-m telescope located at Pico Veleta (Spain) during several sessions, most of them between 2002 and 2008 in the context of a $\lambda$ 3 mm line survey of
IRC~+10216 covering the frequency range 80-116 GHz. The full 3 mm line survey will be published soon elsewhere. The old {\small ABCD} receivers, currently replaced by {\small EMIR} at the 30m telescope, were used. A SIS receiver operating at 3 mm was tuned in single sideband mode, with a typical image rejection larger than 20 dB. Identification of image side band lines was done by shifting the frequency of the local oscillator. Data were taken in the wobbler switching observing mode by nutating the secondary mirror by 3-4$'$ at a rate of 0.5 Hz. Pointing and focus were checked by observing nearby planets and the quasar OJ 287. The back end used was a filterbank with a bandwidth of 512 MHz and a spectral resolution of 1.0 MHz. The intensity scale was calibrated using two absorbers at different temperatures and the atmospheric transmission model ATM
(\cite{cer1985} 1985; \cite{par2001} 2001). We express intensities in terms of $T_A^*$, the antenna temperature corrected for atmospheric absorption and for antenna ohmic and spillover losses. The uncertainty in $T_A^*$ due to calibration is estimated to be around 10 \%. The system temperature ranged from 100 to 150 K. On source integration times ranged from 2 to 20 hours, resulting in $T_A^*$ rms noise levels per 1 MHz channel ranging from less than 1 mK to 3 mK.

\section{Results and discussion}

\begin{table}
\caption{Line parameters observed in IRC~+10216} \label{table:lines}
\centering
\begin{tabular}{lrllr}
\hline \hline
\multicolumn{1}{c}{Transition} & \multicolumn{1}{c}{$\nu_{\rm calc}$} & \multicolumn{1}{c}{$\nu_{\rm obs}$} & \multicolumn{1}{c}{V$_{\rm exp}$} & \multicolumn{1}{c}{$\int T_A^* d$v} \\
                & \multicolumn{1}{c}{(MHz)}                 & \multicolumn{1}{c}{(MHz)}                 & \multicolumn{1}{c}{(km/s)}   & \multicolumn{1}{c}{(K km/s)} \\
\hline
$J'-J''$ & \multicolumn{3}{c}{C$_5$S} \\
\hline
44-43 & 81192.98 & 81193.4(10)$^a$ & 14.5$^b$ & 0.09(3) \\
45-44 & 83038.05 & 83037.8(4) & 13.7(8) & 0.05(1) \\
46-45 & 84883.09 & 84883.5(10) & 16.3(20) & 0.08(2) \\
\hline
$N'_{J'}-N''_{J''}$ & \multicolumn{3}{c}{MgCCH} \\
\hline
9$_{17/2}$-8$_{15/2}$ & 89361.27 & 89361.0(15) & 14.5$^b$ & 0.03(1) \\
9$_{19/2}$-8$_{17/2}$ & 89377.76 & 89378.0(15) & 14.5$^b$ & 0.03(1) \\
10$_{19/2}$-9$_{17/2}$ & 99289.52 & 99288.5(10) & 14.5$^b$ & 0.04(1) \\
10$_{21/2}$-9$_{19/2}$ & 99306.01 & 99305.6(10) & 14.5$^b$ & 0.07(2) \\
\hline
$J'-J''$ & \multicolumn{3}{c}{NCCP} \\
\hline
15-14 & 81131.69 & 81131.9(10) & 14.5$^b$ & 0.03(1) \\
16-15 & 86540.07 & 86539.5(10) & 14.5$^b$ & 0.05(1) \\
18-17 & 97356.60 & 97355.7(10) & 12.4(15) & 0.06(2) \\
\hline
$J'_{K'}-J''_{K''}$ & \multicolumn{3}{c}{SiH$_3$CN$^c$} \\
\hline
9$_0$-8$_0$ & 89509.90 & 89509.5(10) & 16.7(15) & 0.05(1) \\
9$_1$-8$_1$ & 89508.76 &                     &                & \\
10$_0$-9$_0$ & 99454.36 & 99452.5(15)$^a$ & 14.7(15) & 0.08(2) \\
10$_1$-9$_1$ & 99453.08 &                     &                & \\
11$_0$-10$_0$ & 109398.46 & 109397.0(15)$^a$ & 14.5$^b$ & 0.09(3) \\
11$_1$-10$_1$ & 109397.06 &                     &                & \\
\hline
\end{tabular}
\tablenotea{\\
Numbers in parentheses are 1$\sigma$ uncertainties in units of the last digits. $^a$ Line blended. $^b$ Line width has been fixed. $^c$ The $K$=0,1 components of SiH$_3$CN (the strongest ones) are fit to one single line.
}
\end{table}

\subsection{C$_5$S} \label{sec:c5s}

\begin{figure}
\centering
\includegraphics[angle=0,width=\columnwidth]{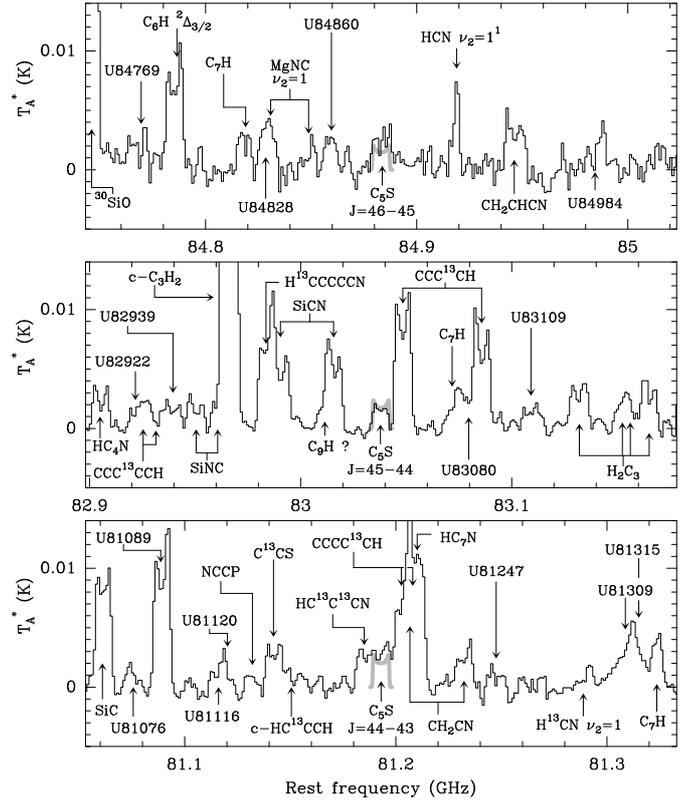}
\caption{Portions of the 3 mm line survey of IRC~+10216 showing the $J=44-43$, $J=45-44$, and $J=46-45$ lines of C$_5$S. The $J=44-43$ line is blended with lines of HC$^{13}$C$^{13}$CN and CCCC$^{13}$CH. The $T_A^*$ rms noise levels per 1 MHz channel are 0.6, 0.5, and 1.0 mK in the bottom, middle, and top panels, respectively. Note the coincidence in frequency of a weak shoulder at the red part of a SiCN line with the $^2\Pi_{3/2}$ $J=201/2-199/2$ line of C$_9$H (middle panel). Thick gray lines are the calculated line profiles under LTE assuming that C$_5$S extends between 5 and 20$''$ with an homogeneous fractional abundance and rotational temperature of 18 K.} \label{fig:c5s}
\end{figure}

The molecule C$_5$S is a relatively heavy linear carbon chain with a $^1\Sigma^+$ ground electronic state, whose rotational spectrum has been recorded in the laboratory in the 5-18 GHz frequency range (\cite{kas1993} 1993; \cite{gor2001} 2001). Its electric dipole moment has been calculated as 4.65 D (\cite{pas1998} 1998; \cite{per2004} 2004). In their observations of IRC~+10216 using the NRAO 140-foot radiotelescope, \cite{bel1993} (1993) reported the detection of a line at 23990.2 MHz which was tentatively
assigned to the $J=13-12$ rotational transition of C$_5$S. To our knowledge, this tentative detection has not been neither confirmed nor refuted by subsequent studies.

Various rotational transitions of C$_5$S with upper level energies $E_{\rm up}$ in the range 90-170 K fall within the frequency coverage of our 3 mm line survey. We have identified at the lower frequency side of the survey three weak emission lines whose frequencies coincide with the $J=44-43$ to $J=46-45$ transitions of C$_5$S. The next $J$ lines are not detected either because they fall in spectral regions observed with an insufficient sensitivity or because they overlap with other strong lines (see Appendix~\ref{app:a}). The line parameters are given in Table~\ref{table:lines} and the lines are
shown in Fig.~\ref{fig:c5s}. The agreement between calculated and observed frequencies is better than 1 MHz, which is good enough taking into account the moderate spectral resolution of the observed spectra (1 MHz) and the non-negligible errors of some hundreds of kHz in the calculated frequencies at 3 mm
(extrapolated from laboratory measurements at cm-wavelengths). In IRC~+10216's spectra the $J=45-44$ and $J=46-45$ lines of C$_5$S appear free of blending with other lines, although the $J=44-43$
line overlaps partially with other weak lines arising from HC$^{13}$C$^{13}$CN and CCCC$^{13}$CH.

Given the limited signal-to-noise ratio of the C$_5$S lines it is
difficult to appreciate a distinct line shape, either U-shaped or
flat-topped, which could permit us to infer whether the size of
the emitting region is smaller or larger than the main beam of the
telescope at these frequencies (29-30$''$). Based on the U-shaped
line profiles observed at 3 mm for the related molecules C$_2$S
and C$_3$S (\cite{cer1987} 1987) we could also expect an extended
emission for C$_5$S. Assuming a source size of radius 20$''$, as
most molecules distributed as a hollow shell in IRC~+10216 (e.g.,
\cite{gue1997} 1997), and adopting the intensities of the three
lines observed at 3 mm and that observed by \cite{bel1993} (1993)
at 24.0 GHz, we derive a column density $N$(C$_5$S) of 1.4
$\times$ 10$^{13}$ cm$^{-2}$ and a rotational temperature $T_{\rm
rot}$ of 18 K. The column density is somewhat lower than that
reported by \cite{bel1993} (1993), mainly because of the lower
source size adopted by these authors. The derived values of
$T_{\rm rot}$ and $N$(C$_5$S) rely to a large extent on the 0.8 cm
line ($E_{\rm up}$ = 8 K) observed by \cite{bel1993} (1993). If we
only consider the lines at 3 mm ($E_{\rm up}$ = 88-96 K) then
$T_{\rm rot}$ needs to be fixed and $N$(C$_5$S) becomes very
sensitive to the adopted choice of $T_{\rm rot}$. For example, for
a rotational temperature of 44 K (as measured for the related
molecule C$_3$S; see Table~\ref{table:colden}) the column density
of C$_5$S decreases by almost one order of magnitude to 2 $\times$
10$^{12}$ cm$^{-2}$. The 3 mm lines involve high energy levels and
have Einstein coefficients for spontaneous emission (and thus
critical densities) about 40 times higher than the 0.8 cm line,
probably implying that they are subthermally excited in the outer
layers of IRC~+10216 and that C$_5$S is not characterized by a
uniform rotational temperature. For $T_{\rm rot}$ = 18-44 K we
find that C$_5$S has a column density in the range (2-14) $\times$
10$^{12}$ cm$^{-2}$. Fig.~\ref{fig:c5s} shows the line profiles
calculated under local thermodynamic equilibrium (LTE) adopting
the parameters of the envelope from \cite{agu2012} (2012) and
assuming that C$_5$S is present in a hollow shell extending from 5
to 20$''$, with an homogeneous fractional abundance and rotational
temperature of 18 K, which may be adequate for the 3 mm lines, as
discussed above. The agreement between modeled and observed line
profiles is fairly good.

The molecule C$_5$S is the largest member of the series of
sulfur-containing carbon chains C$_n$S observed in space. The
shorter members C$_2$S and C$_3$S are observed in various
astronomical regions, although C$_4$S has never been observed. The
column densities derived for these molecules in IRC~+10216 are
given in Table~\ref{table:colden}. In the cases of C$_2$S and
C$_3$S the column densities are quite robust as they result from a
large number of rotational lines observed across the 3, 2, and 1.3
mm bands. That of C$_4$S is just an upper limit while that of
C$_5$S has a large uncertainty, which however could be reduced if
sensitive observations are performed at frequencies around 40 GHz,
where the brightest C$_5$S lines are expected. This spectral
region has been surveyed by \cite{kaw1995} (1995) although no line
attributable to C$_5$S was found, probably due to an insufficient
sensitivity. The carbon chains C$_n$S seem to show a decline in
abundance as the length of the chain increases (albeit the
uncertainty in the value of C$_5$S), as occurs with polyacetylenic
and cyanopolyacetilenic chains (see e.g. \cite{agu2008a} 2008a).
There is no obvious alternation in the abundance between chains
with $n$ odd ($^1\Sigma$ electronic ground state) and those with
an even number of carbon atoms ($^3\Sigma$ electronic ground
state), as may be occurring for the series C$_n$O, of which only
C$_3$O, but not C$_2$O, has been detected in IRC~+10216
(\cite{ten2006} 2006). It is also worth noting that the rotational
temperature of C$_3$S is more than twice that of C$_2$S (see
Table~\ref{table:colden}), which points to differences in the
excitation and may be indicative of radiative pumping to
vibrationally excited states playing an important role in the case
of C$_3$S (see e.g., \cite{agu2008a} 2008a).

\begin{table}
\caption{Rotational temperatures and column densities} \label{table:colden}
\centering
\begin{tabular}{lccr}
\hline \hline
\multicolumn{1}{c}{Molecule} & \multicolumn{1}{c}{$T_{\rm rot}$ (K)} & \multicolumn{1}{c}{$N$ (cm$^{-2}$)}  & Reference \\
\hline
C$_2$S & 18(1)         & 5.0(3) $\times$ 10$^{13}$ & [1] \\
C$_3$S & 44(2)         & 1.7(1) $\times$ 10$^{13}$ & [1] \\
C$_4$S & 20 $^a$     & $<$7 $\times$ 10$^{12}$ $^b$  & [2] \\
C$_5$S & 18-44 & (2-14) $\times$ 10$^{12}$ & [2] \\
MgCCH & 20 $^a$         & 2 $\times$ 10$^{12}$ & [2] \\
NCCP & 20 $^a$        & 7 $\times$ 10$^{11}$ & [2] \\
SiH$_3$CN & 39 $^a$         & 1 $\times$ 10$^{12}$ & [2] \\
\hline
\end{tabular}
\tablenotea{\\
Numbers in parentheses are 1$\sigma$ uncertainties in units of the last digits. A source size of radius 20$''$ has been adopted. $^a$ $T_{\rm rot}$ has been fixed. $^b$ 3$\sigma$ upper limit.\\
References: [1] IRAM 30-m data at $\lambda$ 3, 2, and 1.3 mm. [2]
This study. }
\end{table}

The chemical routes leading to S-bearing carbon chains in IRC~+10216 have been investigated by \cite{mil2001} (2001), who find that the synthesis is probably dominated by reactions of the type
\begin{equation}
{\rm S + C_{2{\tiny \emph{n}}}H \rightarrow C_{2{\tiny \emph{n}}}S + H,} \qquad n = 1, 2, ...
\end{equation}
for S-containing carbon chains with an even number of carbon atoms, while those with an odd number of carbon atoms would form through reactions of the type
\begin{equation}
{\rm CS + C_{2{\tiny \emph{n}}}H \rightarrow C_{2{\tiny \emph{n}}+1}S + H,} \qquad n = 1, 2, ...
\end{equation}
The model of \cite{mil2001} (2001) yields results in reasonable agreement with our observed column densities of C$_n$S chains, although these are somewhat overestimated by the model probably because of the too high initial abundance of CS adopted (4 $\times$ 10$^{-6}$ relative to H$_2$), which may need to be revised down by a factor of 5-6 (\cite{agu2012} 2012).

\subsection{MgCCH} \label{sec:mgc2h}

\begin{figure}
\centering
\includegraphics[angle=0,width=\columnwidth]{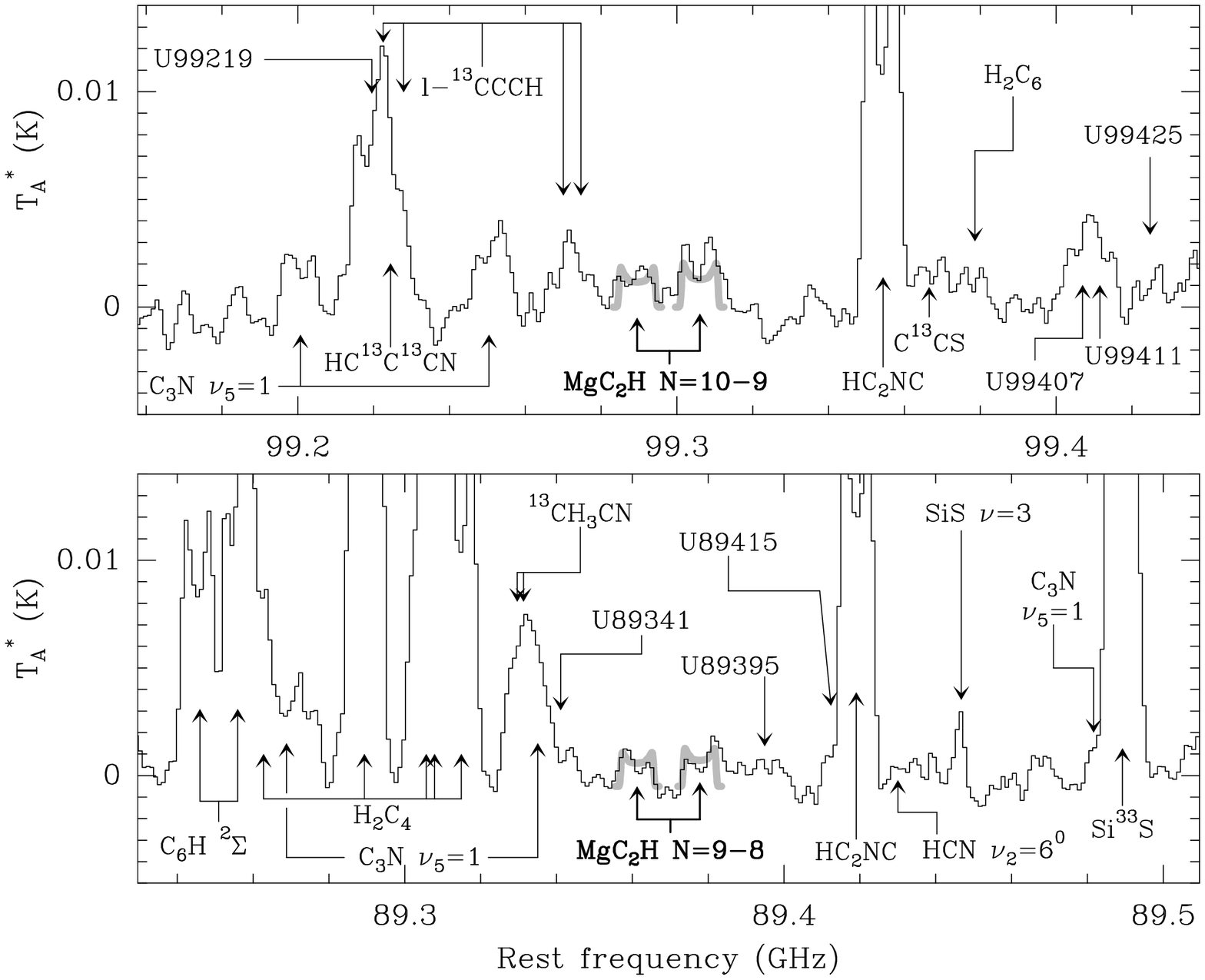}
\caption{Parts of the 3 mm line survey of IRC~+10216 showing the couples of lines tentatively assigned to the $N=9-8$ and $N=10-9$ doublets of MgCCH. The $T_A^*$ rms noise levels per 1 MHz channel are 0.5 and 0.7 mK in the bottom and top panels, respectively. Thick gray lines correspond to the calculated line profiles under LTE assuming that MgCCH extends between 5 and 20$''$ with an homogeneous fractional abundance and rotational temperature of 20 K.} \label{fig:mgc2h}
\end{figure}

A variety of metal cyanides has been observed in the circumstellar
gas of IRC~+10216, among them various involving magnesium such as
the isocyanic and cyanic forms MgNC and MgCN (\cite{gue1986} 1986;
\cite{kaw1993} 1993; \cite{ziu1995} 1995), and more recently the
hydrogenated isocyanic form HMgNC (\cite{cab2013} 2013). Metals
thus show a tendency to combine with cyanide radicals rather than
other types of radicals in IRC~+10216. The ethynyl radical being
as abundant as the cyanide radical in IRC~+10216 it is therefore
reasonable to expect that metal acetylides such as MgCCH could
also be present in the circumstellar gas. Magnesium monoacetylide
(MgCCH) is a linear radical with a $^2\Sigma^+$ electronic ground
state, whose pure rotational spectrum has been characterized in
the laboratory in the spectral range 315-525 GHz (\cite{bre1999}
1999). MgCCH has a moderate electric dipole moment of 1.68 D,
according to the ab initio calculations of \cite{woo1996} (1996),
which indicates that the ethynyl-metal bond probably has a fair
degree of covalent character.

Three rotational transition doublets of MgCCH fall within our 3 mm line survey. The $N=9-8$ and $N=10-9$ at 89.3 and 99.3 GHz, respectively, coincide in frequency with two pairs of non-blended weak lines (see Fig.~\ref{fig:mgc2h}) while the $N=11-10$ at 109.2 GHz falls in a region where the sensitivity of the spectrum is insufficient to detect lines with antenna temperatures of a few mK (see Appendix~\ref{app:a}). The agreement between observed and calculated frequencies is reasonably good, less than 1 MHz (see Table~\ref{table:lines}). It must however be taken into account that the observed lines attributed to MgCCH are weak and the number of lines with similar intensities is relatively large in this region of the spectrum and at this level of sensitivity. The possibility of accidental coincidence cannot be completely ruled out although the assignment to MgCCH is reinforced by the fact that the transitions consist of doublets and both the $N=9-8$ and $N=10-9$ are observed free of blending. Nonetheless, for the moment we prefer to rest cautious and consider the detection of MgCCH as tentative, mainly because of the weakness of the observed lines.

Despite the limited signal-to-noise ratio of the MgCCH lines
observed, these seem to show a U-shaped profile, which would
indicate that the size of the emission is comparable or larger
than the main beam of the IRAM 30-m telescope at the observed
frequencies (25-28$''$). Magnesium isocyanide has been found to be
distributed in the form of a hollow shell of radius 15-20$''$, as
mapped with the IRAM Plateau de Bure interferometer at millimeter
wavelengths (\cite{gue1993} 1993). Based on the U-shaped profiles
and the distribution of the related molecule MgNC we can assume a
source size of radius 20$''$ for MgCCH. Adopting a rotational
temperature of 20 K, of the order of other molecules distributed
in the outer layers of IRC~+10216 (\cite{cer2000} 2000), we derive
a column density $N$(MgCCH) of 2 $\times$ 10$^{12}$ cm$^{-2}$,
that is, $\sim$6 times less than MgNC and $\sim$3 times above
those of MgCN and HMgNC (\cite{cab2013} 2013). As shown in
Fig~\ref{fig:mgc2h}, the MgCCH line profiles calculated under LTE
assuming a rotational temperature of 20 K are in reasonable
agreement with the observations.

The possible chemical routes leading to metal-containing
molecules, including MgCCH, in the outer layers of IRC~+10216 have
been addressed in various studies (\cite{woo1996} 1996;
\cite{pet1996} 1996; \cite{dun2002} 2002; \cite{mil2008} 2008).
Various chemical pathways have been ruled out based on reaction
endothermicities or on the lack of sufficiently abundant
precursors in the circumstellar gas. From these studies it turns
out that the most likely formation route of MgCCH probably follows
closely that proposed by \cite{pet1996} (1996) for the synthesis
of magnesium cyanides, which consists of the radiative association
between Mg$^+$ and large cyanopolyynes followed by the
dissociative recombination of the ionic complex with free
electrons. In the case of MgCCH, the radiative recombination would
take place between Mg$^+$ and large polyynes, leading to
Mg(C$_n$H$_2$)$^+$ complexes, whose recombination with electrons
could yield MgCCH, among other possible fragments. A simple
chemical model similar to that described in \cite{cab2013} (2013)
adopting the rate constants of radiative association between
Mg$^+$ and polyynes calculated by \cite{dun2002} (2002) indicates
that this route is capable of producing MgCCH with an abundance
high enough to account for the column density derived from
observations. An alternative route could be the exchange reaction
between MgNC and CCH to yield MgCCH and CN (the precursor MgNC
being about 6 times more abundant than MgCCH in IRC~+10216),
although theoretical calculations do not conclude whether this
reaction is slightly endothermic or slightly exothermic
(\cite{woo1996} 1996). Other metal acetylides could be present in
IRC~+10216, although they may be difficult to detect if, as found
in the case of Mg, they are less abundant than the corresponding
(iso)cyanides. For example, AlNC is detected in IRC~+10216
(\cite{ziu2002} 2002) but not AlCCH, with $N$(AlCCH)/$N$(AlNC)
$\lesssim$ 22 (\cite{cab2012} 2012). The low dipole moment of
AlCCH ($\sim$0.6 D) makes it difficult to impose a stringent upper
limit on its abundance.

\subsection{NCCP} \label{sec:nccp}

\begin{figure}
\centering
\includegraphics[angle=0,width=\columnwidth]{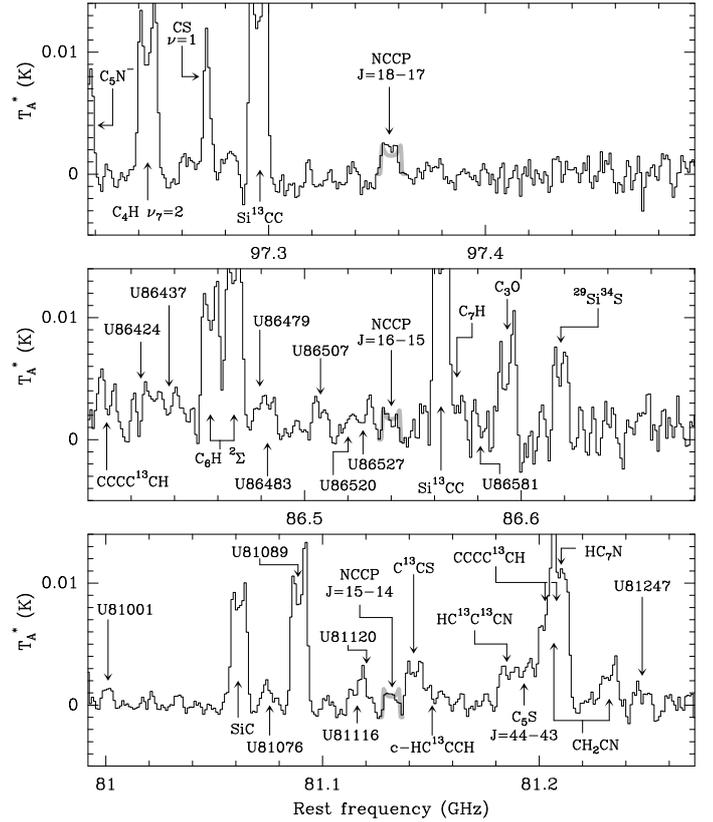}
\caption{Parts of the 3 mm line survey of IRC~+10216 showing the three lines tentatively assigned to the $J=15-14$, $J=16-15$, and $J=18-17$ rotational lines of NCCP. The $J=16-15$ line is marginally detected. The $T_A^*$ rms noise levels per 1 MHz channel are 0.4, 1.2, and 0.8 mK in the bottom, middle, and top panels, respectively. Line profiles calculated under LTE assuming that NCCP extends between 5 and 20$''$ with an homogeneous fractional abundance and rotational temperature of 20 K are shown as thick gray lines.} \label{fig:nccp}
\end{figure}

From a chemical point of view, cyanophosphaethyne (NCCP) can be viewed as a cyanogen molecule (NCCN) in which a nitrogen atom has been substituted by a phosphorus atom. From a spectroscopic point
of view, NCCP is a linear molecule with a closed-shell singlet electronic ground state. Its pure rotational spectrum has been characterized in the laboratory over a broad spectral range, from 25 to 820 GHz (\cite{biz2001} 2001). Having a moderately large dipole moment of 3.44 D, as measured by \cite{coo1980} (1980), it has been suggested that this molecule could be a good candidate for detection in astronomical sources such as IRC~+10216 (\cite{biz2001} 2001), where other P-bearing molecules such as CP,
PN, HCP, PH$_3$, and CCP have been already detected (\cite{gue1990} 1990, 2000; \cite{agu2007}, 2007, 2008b, 2014; \cite{milam2008} 2008; \cite{ten2008} 2008; \cite{hal2008} 2008).

Seven rotational lines of NCCP from $J=15-14$ to $J=21-20$, with upper level energies in the range 30-60 K, fall within the 3 mm line survey. A weak line with $T_A^*$ $\sim$2-3 mK lying at 97355.7 $\pm$ 1 MHz coincides in frequency, within less than 1 MHz, with the $J=18-17$ transition of NCCP (see Table~\ref{table:lines} and Fig.~\ref{fig:nccp}). To confirm whether this assignment is correct we have searched for other rotational transitions of NCCP within the 3 mm band. A weak line at 81131.9 $\pm$ 1 MHz, lying in one of the most sensitive regions of the 3 mm spectrum, might be assigned to the $J=15-14$
transition while the $J=16-15$ can be identified with a marginal line at 86539.5 $\pm$ 1 MHz. The rest of transitions of NCCP within the 3 mm band either fall in regions where the sensitivity achieved is insufficient to detect such weak lines or are affected by severe blendings with other strong lines (see Appendix~\ref{app:a}). The agreement between calculated and observed frequencies is reasonably good and, within the sensitivity reached, there are not missing lines in the 3 mm line survey. On the other hand, line confusion and accidental coincidence may be an issue at this level of sensitivity in the 3 mm spectra of IRC~+10216. We must therefore be cautious and consider the identification of NCCP as tentative.

Given the limited signal-to-noise ratio of the three lines assigned to NCCP it is difficult to distinguish whether the line profiles are more U-shaped or flat-topped. Assuming that this molecule is formed in the outer layers, as seems to be the case of the related radical CCP (\cite{hal2008} 2008), we may adopt for
NCCP a distribution size with a radius of 20$''$ and a rotational temperature of 20 K, typical parameters for molecules distributed in the outer shells. The column density of NCCP derived is then 7 $\times$ 10$^{11}$ cm$^{-2}$, about 50-100 times less abundant than CP and slightly less abundant than CCP. The column densities of CP and CCP in IRC~+10216 are in the ranges 3.9-8 $\times$ 10$^{13}$ and 1.2-2.9 $\times$ 10$^{12}$ cm$^{-2}$, respectively, according to \cite{gue1990} (1990), \cite{hal2008} (2008), and an
analysis of our IRAM 30-m data. As can be seen in Fig~\ref{fig:nccp}, the line profiles of NCCP calculated under LTE assuming a rotational temperature of 20 K show a good agreement with the observed ones.

The chemistry of phosphorus-containing molecules in circumstellar envelopes such as IRC +10216 has been discussed by \cite{mac2001} (2001) and \cite{agu2007} (2007), although the formation of NCCP
has not yet been addressed. Little is known on the chemical kinetics of this species. We can imagine that formation routes similar to those yielding cyanogen (NCCN), which are better studied, could also work for NCCP. The formation of NCCN in the reaction of HCN and CN does not occur at low temperatures because
of an activation barrier (\cite{yan1992} 1992), although the reaction between HNC and CN may provide an efficient formation route to NCCN as long as the reaction is barrierless and both HNC and CN are abundant in IRC~+10216 (\cite{pet2003} 2003). Unfortunately, NCCN does not have permanent dipole moment and cannot be detected through its rotational spectrum. By analogy with NCCN, we can imagine that NCCP might be efficiently formed in IRC~+10216 through the reaction
\begin{equation}
{\rm HNC + CP \rightarrow NCCP + H,}
\end{equation}
which is slightly exothermic, by $\sim$15 kcal mol$^{-1}$ adopting the formation enthalpy of NCCP from \cite{pha2001} (2001), and whose reactants are abundant enough in IRC~+10216.

\subsection{SiH$_3$CN}

\begin{figure}
\centering
\includegraphics[angle=0,width=\columnwidth]{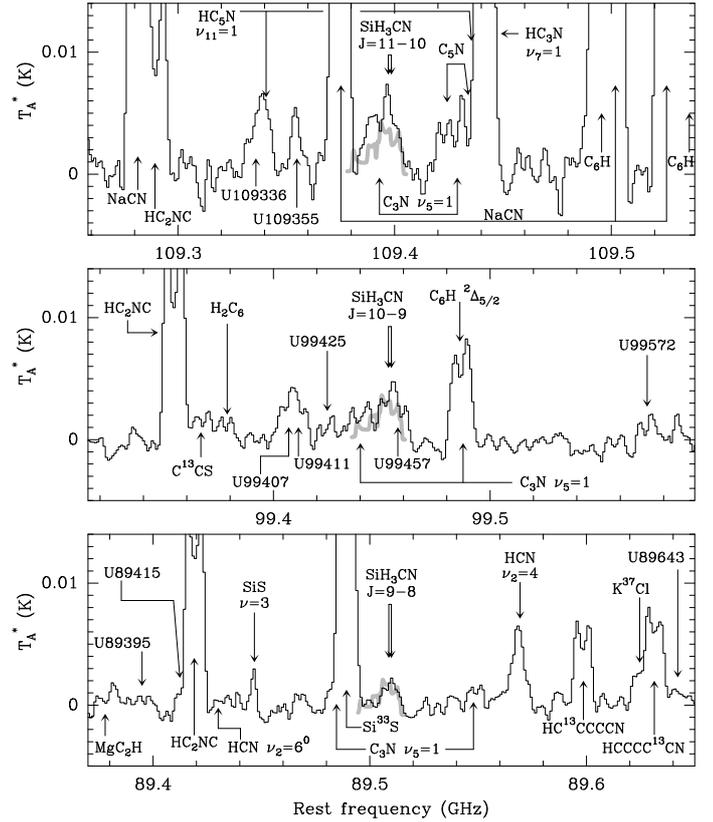}
\caption{Portions of the 3 mm line survey of IRC~+10216 showing the $J=9-8$ to $J=11-10$ rotational lines of SiH$_3$CN. Each $J$ line is composed of a set of $K$ components whose line profiles overlap. The $K=0$ and 1 components, being the strongest, are indicated by arrows. The $T_A^*$ rms noise levels per 1 MHz channel are 0.5, 0.6, and 1.2 mK in the bottom, middle, and top panels, respectively. The $J=10-9$ line is blended with a weak unidentified line at the blue side and both the $J=10-9$ and $J=11-10$ lines are blended at the red side with a weak component of C$_5$N in its $\nu_5=1$ vibrational state. These blended lines cannot by themselves explain the whole emission features centered at the frequencies of SiH$_3$CN. Thick gray lines correspond to the calculated line profiles under LTE assuming that SiH$_3$CN extends between 5 and 20$''$ with an homogeneous fractional abundance and rotational temperature of 39 K.} \label{fig:sih3cn}
\end{figure}

Silyl cyanide, SiH$_3$CN, is a prolate rotor with a closed electronic shell. Its rotational spectrum has been recorded in the laboratory at microwave and millimeter wavelengths and its electric dipole moment has been experimentally determined to be as large as 3.44 D (\cite{pri1998} 1998).

Three rotational transitions of SiH$_3$CN fall within the 3 mm line survey, $J=9-8$ to $J=11-10$, and the three can be identified with weak emission features in the observed spectrum. Because lines are relatively broad in IRC~+10216, $\sim$29 km s$^{-1}$ or $\sim$10 MHz at 3 mm, each transition of SiH$_3$CN is expected to show a complex profile resulting from the overlap of the different $K$ components, the $K=0$ and 1 being the strongest ones. Fig.~\ref{fig:sih3cn} shows the SiH$_3$CN lines while Table~\ref{table:lines} gives the line parameters. Only the positions of the $K=0$ and 1 components are indicated. The $J=9-8$ line appears free of blend with other lines, although the $J=10-9$ and $J=11-10$ overlap partially at the red side with a weak line of C$_5$N in its $\nu_5=1$ vibrational state. Moreover, in the case of the $J=10-9$ line the emission feature observed implies that there must be a weak overlapping unidentified line at $\sim$99457 MHz. The computation of the line profiles expected for SiH$_3$CN helps to bring some light into the interpretation of the observed emission features. Assuming that SiH$_3$CN is distributed as a hollow shell of radius between 5 and 20$''$ with an homogeneous rotational temperature of 39 K, that derived for the related molecule CH$_3$CN according to our IRAM 30-m data, a LTE excitation calculation yields the line profiles shown in Fig.~\ref{fig:sih3cn}. It is seen that the line profiles of SiH$_3$CN are dominated by the $K=0$ and 1 components with minor contributions from higher $K$ components at lower frequencies. The observed emission features cannot be fully explained by the C$_5$N $\nu_5=1$ lines and the unidentified line at 99457 MHz, and thus are consistent with the presence of SiH$_3$CN. However, because two of the three observed transitions of SiH$_3$CN are partially blended with other lines and because of the limited signal-to-noise ratio achieved, for the moment we consider the identification of SiH$_3$CN in IRC~+10216 as tentative.

A relatively low column density of 10$^{12}$ cm$^{-2}$ is derived for SiH$_3$CN adopting a rotational temperature of 39 K and a source size of radius 20$''$, which makes SiH$_3$CN the least abundant Si-containing molecule detected to date in IRC~+10216. Its column density is comparable to, although somewhat below, that of SiCN and SiNC ($\sim$2 $\times$ 10$^{12}$ cm$^{-2}$; \cite{gue2000} 2000, 2004). The synthesis of SiH$_3$CN is quite uncertain, although the analogous species CH$_3$CN has been better studied. A quite direct synthetic pathway to CH$_3$CN would be the reaction between CH$_4$ and CN although it has an activation barrier and thus becomes too slow at low temperatures (\cite{sim1993} 1993). In the absence of better constraints we could expect that the route to SiH$_3$CN from SiH$_4$ and CN is also probably closed at low temperatures. A more efficient route to CH$_3$CN in IRC~+10216 involves the ion CH$_3$CNH$^+$, whose dissociative recombination with electrons yields both CH$_3$CN and CH$_2$CN (\cite{agu2008a} 2008a). In the case of SiH$_3$CN, a similar route involving the ion SiH$_3$CNH$^+$ could also work, although the details on the chemical kinetics of the reactions involved are yet to be investigated.

\section{Summary}

We have detected various weak lines in the course of a IRAM 30-m $\lambda$ 3 mm line survey of IRC~+10216, three of which are identified with rotational transitions of C$_5$S, confirming a previous tentative detection of this molecule by \cite{bel1993} (1993) based on a line at 0.8 cm, while some others are tentatively assigned to rotational transitions of three new molecules not yet observed in space. These three molecules are MgCCH, the first metal acetylide detected in space, and NCCP and SiH$_3$CN, two exotic molecules which are the phosphorus and silicon analogs of cyanogen (NCCN) and methyl cyanide (CH$_3$CN).

The S-bearing carbon chain C$_5$S is likely to have a non uniform rotational temperature, with values in the range 18-44 K, which translates to column densities in the range (2-14) $\times$ 10$^{12}$ cm$^{-2}$, implying that C$_5$S is less abundant than C$_3$S in the envelope of IRC~+10216. For MgCCH, NCCP, and SiH$_3$CN we derive column densities of the order of 10$^{12}$ cm$^{-2}$, which implies that MgCCH has an abundance in between that of MgNC and those of MgCN and HMgNC, while NCCP and SiH$_3$CN become the least abundant P- and Si-bearing molecules observed to date in IRC~+10216.

Based on the behavior of similar molecules it is likely that the four molecules are formed in the outer circumstellar layers of IRC~+10216. The synthesis of C$_5$S probably occurs in a similar fashion to other sulfur-containing carbon chains such as C$_3$S, while that of MgCCH probably takes place through chemical routes similar to those proposed for the formation of metal cyanides. The formation of NCCP and SiH$_3$CN is more uncertain and is suggested to occur through analogous pathways to those leading to NCCN and CH$_3$CN.

The identification of these four molecules through such weak lines suggests that pushing observations towards very low noise levels will probably bring a good number of molecular discoveries in chemically rich environments such as IRC~+10216.

\begin{acknowledgements}

M.A. and J.C. thank Spanish MINECO for funding support through grants CSD2009-00038, AYA2009-07304, and AYA2012-32032. We thank the anonymous referee for a critical and constructive report.

\end{acknowledgements}

\clearpage

\appendix

\section{Spectra covering detected and non detected lines in the 3 mm band} \label{app:a}

In the case of silyl cyanide the three rotational transitions covered within the 3 mm band are observed, but for the molecules C$_5$S, MgCCH, and NCCP not all rotational transitions lying in the 3 mm band (80-116 GHz) are detected. The lines which are not detected either fall in spectral regions where the sensitivity achieved is not good enough or are blended with stronger lines of other molecules.

\begin{figure}
\centering
\includegraphics[angle=0,width=\columnwidth]{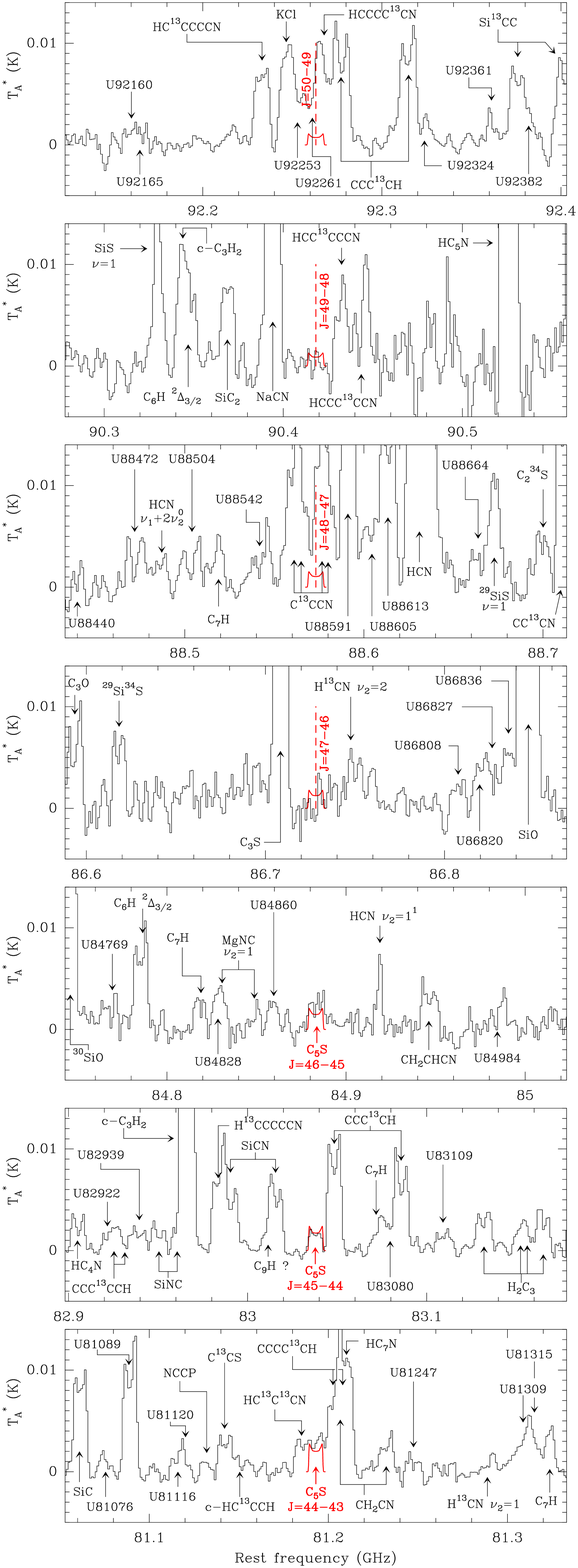}
\caption{Spectra covering the C$_5$S lines in the 80-93 GHz frequency range. Detected lines are indicated by red arrows and non detected lines by red dashed lines. LTE calculated line profiles are shown in red.} \label{fig:app-c5s}
\end{figure}

In Fig.~\ref{fig:app-c5s} we show the spectra covering the $J=44-43$ to $J=50-49$ rotational transitions of C$_5$S, lying in the low frequency side (80-93 GHz) of the 3 mm band. Only the three rotational transitions lying at the lowest frequencies ($J=44-43$, $J=45-44$, and $J=46-45$) are detected (lines also shown in Fig.~\ref{fig:c5s} and already discussed in section~\ref{sec:c5s}). The spectral regions where the next higher $J$ lines of C$_5$S are expected are shown in the upper panels of Fig.~\ref{fig:app-c5s}. None of these transitions can be identified with clear emission features in the observed spectra. The spectral region around the $J=47-46$ and $J=49-48$ transitions are relatively noisy. The antenna temperature rms noise level per 1 MHz channel is 1.5 mK around the $J=47-46$ transition, while the expected C$_5$S line intensity is similar, making very difficult to ascertain whether or not the weak C$_5$S line is present. The situation is even worse for the $J=49-48$ transition, with an expected antenna temperature of just $\sim$1 mK and a rms noise level of 1.8 mK per 1 MHz channel. The $J=48-47$ and $J=50-49$ transitions of C$_5$S are severely blended with other lines much stronger. At higher frequencies the chances of observing a C$_5$S line become lower because the expected C$_5$S line intensities decrease as $J$ increases (C$_5$S is a quite heavy rotor and levels with $J>50$ become poorly populated at rotational temperatures around 20 K) and also because in our 3 mm line survey of IRC~+10216 spectra usually become less sensitive with increasing frequency.

\begin{figure}
\centering
\includegraphics[angle=0,width=\columnwidth]{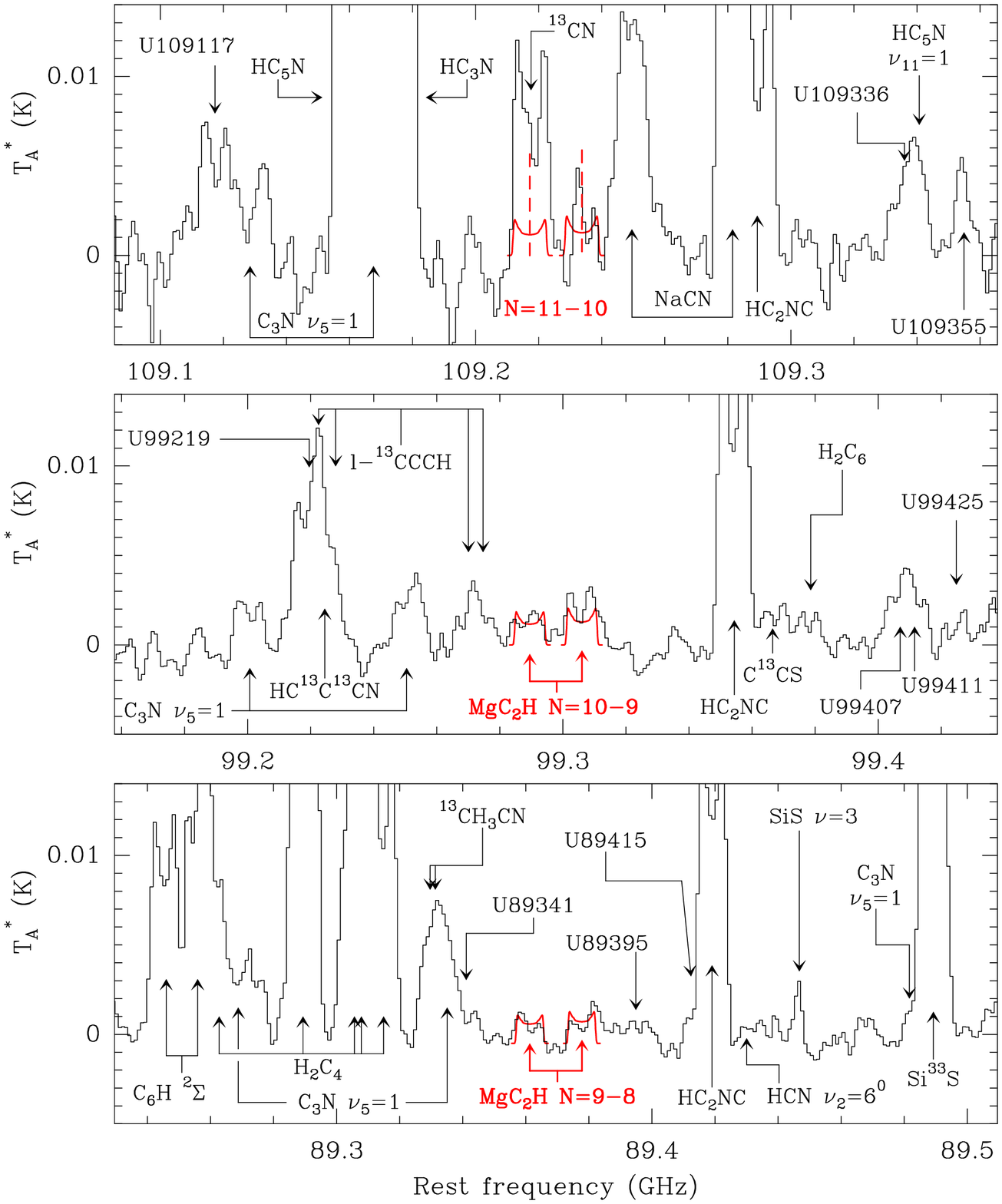}
\caption{Spectra covering the MgCCH lines in the 3 mm band. Detected lines are indicated by red arrows and non detected lines by red dashed lines. LTE calculated line profiles are shown in red.} \label{fig:app-mgc2h}
\end{figure}

As concerns the molecule MgCCH, there are three doublets of rotational transitions lying within the 3 mm band (see Fig.~\ref{fig:app-mgc2h}). The two doublets lying at lower frequencies are identified with weak emission features (see also Fig.~\ref{fig:mgc2h} and section~\ref{sec:mgc2h}) but the one lying at 109.2 GHz, corresponding to the $N=11-10$ rotational transition, cannot be clearly distinguised in the 3 mm line survey data. The low frequency component of this doublet is completely blended with a stronger line arising from an hyperfine component of the $N_J=1_2-0_1$ rotational transition of $^{13}$CN, while the high frequency component cannot be appreciated due to an insufficient sensitivity (the $T_A^*$ rms noise level is 1.8 mK per 1 MHz in this spectral region and the MgCCH component is expected with an intensity of just $\sim$1.5 mK).

\begin{figure}
\centering
\includegraphics[angle=0,width=\columnwidth]{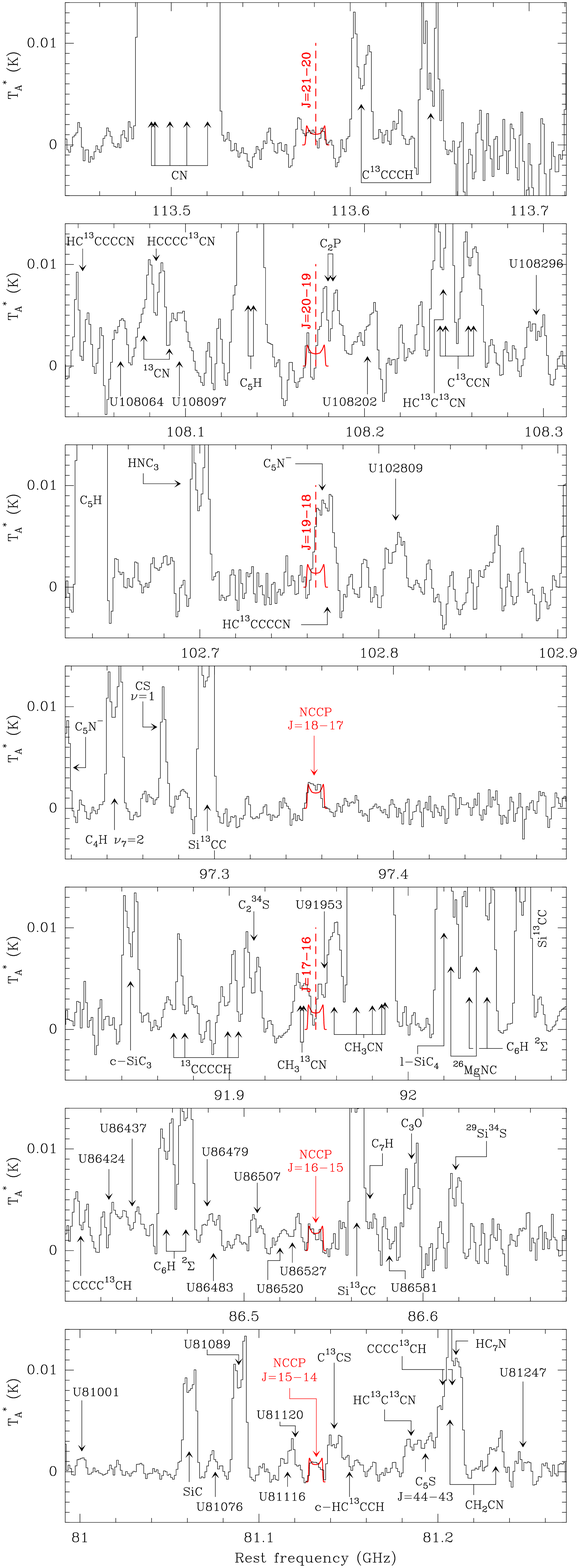}
\caption{Spectra covering the NCCP lines in the 3 mm band. Detected lines are indicated by red arrows and non detected lines by red dashed lines. LTE calculated line profiles are shown in red.} \label{fig:app-nccp}
\end{figure}

The linear rotor NCCP has also several rotational transitions lying in the 80-116 GHz range (see Fig.~\ref{fig:app-nccp}), from which only three ($J=15-14$, $J=16-15$, and $J=18-17$) are identified with weak emission features (see also Fig.~\ref{fig:nccp} and section~\ref{sec:nccp}). The $J=17-16$ transition falls in a spectral region crowded by stronger lines arising from CH$^{13}$CN, and unidentified line at 91953 MHz, and CH$_3$CN, making very difficult to infer its presence. The $J=19-18$ transition is partially blended with a stronger line corresponding to C$_5$N$^-$, and in any case the limited sensitivity of the spectrum (rms of 1.5 mK per 1 MHz) makes it very difficult to distinguish the NCCP line from the noise. An even worse situation occurs in the case of the $J=20-19$ transition, which falls in a spectral region where it partially overlaps with a line of C$_2$P and the data is quite noisy (rms of 2.1 mK per 1 MHz). The last transition of NCCP within the 3 mm band is the $J=21-20$, which falls in a spectral region where the noise level is moderately low (rms of 1.1 mK per 1 MHz if we do not consider the much noiser region at frequencies higher than 113600 MHz), although it is not low enough to allow for a clear detection of lines with antenna temperatures of $\sim$1 mK, as expected for this NCCP line.

\end{document}